# Complete Au-free Metallization Platform for UWBG $Al_{0.65}Ga_{0.35}N$ channel MOS-HEMTs using Ti/Al/Ti Ohmic Contacts and Refractory TiN/Pt Gates

Mafruda Rahman[1], Dharun Manoharan[1], Tariq Jamil[1], Abdullah Al Mamun Mazumder[1], Muhammad Ali[1], Anthony Joyce[1], Chandan Joishi[2], James G. Fiorenza[2], Manoj Kumar Chandra Mohan[1], Grigory Simin[1], and Asif Khan[1]

[1]Department of Electrical Engineering, University of South Carolina, Columbia, SC - 29208, USA

[2]Analog Devices Incorporated, Boston, MA - 02110, USA

**Authors to whom correspondence should be addressed:** mafruda@email.sc.edu

Au-free ohmic contacts and refractory gate stacks are critical for integrating ultra-wide bandgap (UWBG) Al-rich AlGaN transistors with Si-CMOS technology. In this work, we demonstrate, to our knowledge, the first complete Au-free metallization platform for $Al_{0.65}Ga_{0.35}N$-channel MOS-HEMTs, combining electron-beam-evaporated and sputtered Ti/Al/Ti ohmic contacts with sputtered TiN and Pt gate metals. The Ti/Al/Ti contacts achieved a contact resistance of ~7.5–7.7 Ω·mm while being fully Au-free. This contact resistance is achieved using a 600°C anneal, substantially below the 850–950°C typically required for Au-capped Ti/Al/X/Au contacts on Al-rich AlGaN, indicating a reduced thermal budget and improved compatibility with Silicon-CMOS foundries. TiN gates deliver drain-current density up to 450 mA/mm at $V_{th}$ = −13 V, whereas Pt gates shift $V_{th}$ to −2.5 V with a reduced current drive of 220 mA/mm, reflecting the effects of gate work function. TiN-gated devices exhibit lower current degradation under pulsed operation and a higher critical breakdown field ($E_c \approx 3.5$ MV/cm) than Pt-gated devices ($E_c \approx 2.98$ MV/cm), measured without field-plating at $L_{GD}$ = 2.3 µm. Collectively, these results establish a low-thermal-budget, fully Au-free, CMOS-compatible metallization platform for high-performance Al-rich AlGaN UWBG power electronics.

Ultra-wide bandgap (UWBG) $Al_xGa_{1-x}N$ ($x < 0.6$), and extreme-bandgap (EBG; $x > 0.6$) materials and devices have emerged as a promising platform for next - generation high-power switching and RF electronics [1], [2], [3], [4], [5]. Their high critical breakdown field, high electron-saturation velocities, and robust polarization-induced 2DEG channel give Al-rich AlGaN/GaN and $Al_xGa_{1-x}N/Al_yGa_{1-y}N$ ($y-x \geq 0.2$) HEMTs substantial advantages over Si and SiC power devices [6], [7]. While AlGaN/GaN power devices are commercialized, scalable manufacturing of emerging Al-rich AlGaN UWBG/EBG HEMTs requires processes compatible with high-volume Si-CMOS foundries [8], [9], [10].

A major barrier to monolithic III-nitride/Si-CMOS integration is reliance on Au-based ohmic metallization. Conventional Ti/Al/X/Au (X = Ni, Ti, Mo, Pt) stacks require rapid thermal annealing at 850–950°C [1] - [5], [11], [12], [13]. Annealing enables Ti-nitride reactions that facilitate tunneling [14], [15], while the Au cap resists oxidation and improves stability [12]. However, high-temperature processing oxidizes the AlGaN surface and generates nitrogen-vacancy defects that promotes field-emission (FE) tunneling [15], [16] yet degrade dynamic performance and reliability [14], [17], while increasing gate to source/drain shorting risk through metal protrusions and poor edge definition in high-frequency HEMTs [18].

Au is incompatible with Si-CMOS due to its diffusivity in silicon [17], [19] besides Au increases both cost and thermal budget. Consequently, there is an important need to transition to Au-free ohmic-contact architectures, such as Ti/Al/Ti or Ti/Al/Ni, that are compatible with Si-CMOS foundry processes. Au-free stacks like Ti/Al/Ti or Ti/Al/Ni have succeeded for GaN-channel HEMTs [15],[20],[21],[22] though remain unexplored for UWBG AlGaN-channel HEMTs, whereas wider bandgap and higher barrier height complicate ohmic formation. UWBG $Al_xGa_{1-x}N$ HEMTs thus rely on reverse-graded $n^{++}$ contact layers [1] - [5], [11].

Using conventional Au-based metallization, several groups have reported contact resistances in the range of 1.4 – 4.7 Ω·mm [1] - [5], [11], [23]. Ni/Au or Pt/Ti/Au gate stacks offer favorable Schottky characteristics and low gate resistance [1] - [5], [11]; however, the incorporation of Au prevents compatibility with Si-CMOS fabrication lines. TiN offers high work function, thermal stability, and reduced trapping [24], [25]. Similarly, Pt has been investigated as a refractory high-work-function gate material because of its excellent chemical inertness and thermal stability [26]. Though W/Pd [27] and TiN [28], [29] Schottky gates explored in Al-rich AlGaN-channel HEMTs, source/drain contacts remained Au-based, leaving completely Au-free UWBG metallization unrealized.

In this work, a fully Au-free metallization platform for EBG Al-rich AlGaN MOS-HEMT is demonstrated by integrating low-thermal-budget Ti/Al/Ti source/drain ohmic contacts with sputtered refractory TiN and Pt gate contacts. The effects of ohmic-contact deposition method (electron-beam evaporation vs. sputtering) and gate-metal selection on contact resistance, threshold voltage, current drive, dynamic performance, and on-state resistance are systematically assessed. The Au-free Ti/Al/Ti stack achieves low-resistance ohmic contacts following annealing at only 600°C, while refractory TiN and Pt gates offer complementary threshold-voltage control and dynamic characteristics. Collectively, these results establish a practical CMOS-process-compatible metallization strategy for next-generation EBG Al-rich AlGaN power devices. Such an Au-free platform is critical for integrating AlGaN-channel UWBG HEMTs into standard Si-CMOS process lines, where Au contamination is restricted.

The HEMT heterostructure was grown by metal-organic chemical vapor deposition (MOCVD) on a sapphire substrate with a 2.6-μm-thick AlN buffer. As illustrated in Fig. 1(a). The epitaxial structure consists of a 140-nm unintentionally doped (UID) $Al_{0.85}Ga_{0.15}N$ back-barrier layer, followed by a 100-nm UID $Al_{0.65}Ga_{0.35}N$ channel layer. A polarization-engineered AlN spacer was introduced to enhance two-dimensional electron gas (2DEG) confinement, followed by a 25-nm $Al_{0.85}Ga_{0.15}N$ barrier layer. A highly Si-doped 5-nm $Al_{0.20}Ga_{0.80}N$ cap layer was employed on 100-nm reverse-composition-graded (RGL) $n^{++}$-$Al_{0.85\rightarrow0.20}Ga_{0.15\rightarrow0.80}N$ contact layers to compensate the polarization doping and facilitate low-resistance ohmic-contact formation.

The device fabrication starts with deposition of Au-free Ti/Al/Ti (50/70/30 nm) source/drain ohmic contacts by either electron-beam evaporation or DC sputtering, followed by RTA at 600°C in a nitrogen ambience to form ohmic contacts. The RGL layer in the access regions was subsequently removed by inductively coupled plasma reactive-ion etching (ICP-RIE), followed by mesa isolation to define the active device region. A 10-nm $SiO_2$ gate dielectric was deposited by plasma-enhanced chemical vapor deposition (PECVD), after which 400-nm-thick refractory TiN or Pt gate contacts were deposited by DC sputtering and patterned to complete MOS-HEMT fabrication. This work is evaluated at 600°C annealing conditions; systematic temperature sweeps to identify the minimum viable annealing temperature and further reduce contact resistance are left for future work.

Four metallization schemes were implemented by combining electron-beam-evaporated or sputtered Ti/Al/Ti ohmic contacts with sputtered TiN or Pt gate contacts to investigate the influence of metallization on device performance. These devices are denoted as Type A (electron-beam Ti/Al/Ti + TiN), Type B (electron-beam Ti/Al/Ti + Pt), Type C (sputtered Ti/Al/Ti + TiN), and Type D (sputtered Ti/Al/Ti + Pt), as summarized in Table I. For comparison, the performance of the Au-free devices (this work) is shown with the Au-based contact previously reported from our group [1] on Al-rich AlGaN MOS-HEMT also listed in Table I. All electrical measurements were performed on devices with a gate length of 1.7 μm, a gate-to-source spacing of 2.0 μm, and a gate-to-drain spacing of 2.3 μm.

The contact resistance ($R_c$) and sheet resistance ($R_{sh}$) were extracted using the standard transfer-length method (TLM), as shown in Fig. 1(b) and (c). The insets show linear current–voltage (I–V) characteristics for TLM patterns with contact spacings ranging from 4 to 20 μm, confirming ohmic behavior. The extracted contact resistances are 7.5 Ω·mm (fit value: 0.995) and 7.7 Ω·mm (fit value: 0.974) for the electron-beam-evaporated and sputtered Ti/Al/Ti contacts, indicating comparable contact formation. However, the electron-beam-evaporated contacts exhibit a lower sheet resistance of 1830 Ω/□ than the sputtered contacts, 2802 Ω/□. This difference may be attributable to the gentle, low-kinetic-energy deposition associated with electron-beam evaporation, which can avoid high-energy ion bombardment and plasma-induced defect generation in the underlying RGL layer, thereby preserving carrier mobility and reducing carrier compensation [24].

Fig. 2(a)–(d) presents the DC and pulsed output characteristics at ($V_{GS,Q}$, $V_{DS,Q}$) = (0 V,0 V) condition of the four Au-free Al-rich AlGaN MOS-HEMTs listed in Table I. Pulsed I–V characteristics were measured using a Keithly 4200A-SCS Parameter Analyzer with a pulse width of 6 μs and a 1 ms off-time (duty cycle ≈ 0.6%). The electron-beam-evaporated Ti/Al/Ti ohmic contacts combined with a TiN gate (Type A) exhibit the highest drain-current density, 450 mA/mm, and the lowest DC on-resistance, 28 Ω·mm. Replacing the TiN gate with Pt (Type B) reduces the maximum drain current to 283 mA/mm and increases the on-resistance to 33 Ω·mm. Devices with sputtered Ti/Al/Ti ohmic contacts (Types C and D) show lower current drive, with maximum drain currents of 278 mA/mm and 220 mA/mm and on-resistances of 49 Ω·mm and 41 Ω·mm.

Fig. 3(a) compares transfer characteristics of the four-type Au-free devices at $V_{DS}$ = 20 V, revealing a systematic threshold-voltage ($V_{th}$) shift with gate material and ohmic-contact method. TiN-gated devices show more negative $V_{th}$ -13 V (Type A), -10V (Type C), while Pt-gated devices shift less-negative −8 V (Type B), −2.5 V (Type D). According to Tapajna et al.'s analytical threshold-voltage model [30], $V_{th}$ in-GaN MOS-HEMTs are governed by the combined effects of gate-metal work function, polarization charge, oxide work function increases the effective Schottky barrier and gate potential, resulting in a less-negative threshold voltage agreeing with Lo et.al. [31]. Fig. 3(b) compares gate-leakage ($I_G$-$V_G$) characteristics of the four MOS-HEMTs at $V_{DS}$ = 20V, with leakage currents ranging from $10^{-8}$ to $10^{-6}$ A/mm.

Figure 4 (a-d) shows Pulsed I-V characteristics at a pulse width of 6 μs and 1 ms off-time (duty cycle ≈ 0.6%) under different quiescent bias conditions of ($V_{GS,Q}$, $V_{DS,Q}$) = (0,0) V, (−20, 20) V, and (-20, 40) V. The measured drain current degradation was 4.2% to 6.8% for Type A, 19.4% to 21.9% for Type B, 0.8% to 6.1% for Type C, and 10.5% to 18.2% for Type D as the quiescent bias increases from (0,0) V to (-20, 40) V. Overall, TiN-gated devices (Type A and C) exhibit lower current collapse than Pt-gated devices, consistent with reports on ALD TiN Schottky-gated AlGaN/GaN HEMTs [32] and the TiN ability to mitigate trapping and improve dynamic stability.

Fig. 5 shows dynamic $R_{on}$ of all Au-free device type (A, B, C and D) measured at pulse width of 6 μs and 1 ms off-time (duty cycle ≈ 0.6%) under quiescent biases for ($V_{GS,Q}$,$V_{DS,Q}$) = (0,0), (−20, 20) and (−20, 40) V. For all devices, $R_{on}$ increases with increasing $V_{DS,Q}$, indicating the influence of charge trapping under high-electric-field stress. Type A exhibits the smallest degradation, ranging from ~27 to ~30 Ω·mm under different quiescent biases. In contrast, Types B and C show more pronounced degradation, reaching approximately 72 Ω·mm and 71 Ω·mm at ($V_{GS,Q}$, $V_{DS,Q}$) = (−20, 40) V. Type D exhibits an intermediate increase from 45 Ω·mm to 51 Ω·mm. The superior dynamic $R_{on}$ stability of Type A is consistent with its lower sheet resistance, higher drain current, and reduced current collapse, indicating that the combination of electron-beam Ti/Al/Ti ohmic contacts and a TiN gate electrode provides effective carrier transport under dynamic operation.

We also investigated the off-state breakdown characteristics of TiN and Pt-gated non-field-plated devices shown in Fig. 6 and 7. Fig. 6 represents the off-state $I_D$-$V_D$ and $I_G$-$V_D$ characteristics of TiN-gated device with $L_{GD}$ ≈ 2.3 μm at $V_{GS}$ = -30V. These devices were submerged in Fluorinert FC-70 to prevent atmospheric breakdown during the high-voltage measurement. The TiN-gated device exhibits a breakdown voltage of ~ 805 V for $L_{GD}$ ≈ 2.3 μm, corresponding to an average electric field of ≈ 3.5 MV/cm.

In Fig. 7, the off-state $I_D$-$V_D$ and $I_G$-$V_D$ characteristics of Pt-gated device device with $L_{GD}$ ≈ 2.3 μm at $V_{GS}$ = -30V is shown. The device exhibited a breakdown voltage of ~ 686 V, corresponding to an average electric field of ≈ 2.98 MV/cm for $L_{GD}$ ≈ 2.3 μm. TiN and Pt gated devices exhibit similar gate leakage at $V_{GS}$ = -30V.The variation in breakdown voltage with TiN and Pt gate metals suggests that breakdown is not only influenced by surface leakage, also by field distribution, gate-metal properties and device-to-device variation.

Fig. 8 illustrates the average electric field across different maximum drain current densities for TiN and Pt-gated devices in comparison with other Au-gated devices for Al-rich AlGaN HEMTs. The TiN-gated device achieved a maximum current density of ~450 mA/mm and $E_c$ ~ 3.5 MV/cm, whereas Pt gated device showed 283 mA/mm and $E_c$ ~ 2.98 MV/cm for 2.3 µm device. The Au- free refractory gated devices show comparable behavior in terms of maximum drain current density and average electric field with the reported Au-gated devices [1], [2], [33], [34], [35], [36], [37], [38], [39], [40], [41], [42], [43], [44], [45].

A fully Au-free, CMOS-compatible metallization platform is demonstrated using Ti/Al/Ti ohmic contacts with TiN/Pt gates. Ti/Al/Ti exhibits a contact resistance of ~7.5–7.7 Ω·mm at reduced 600 °C annealing temperature. Electron-beam contacts outperformed sputtered contacts in conduction and $R_{on}$, whereas the TiN gate exhibits a comparatively higher current drive, while the Pt gate demonstrates a $V_{th}$ shift toward the low-negative region. The electron-beam Ti/Al/Ti-TiN combination performed best overall performance with a maximum drain current of 450 mA/mm, a DC on-resistance of 28 Ω·mm, and a threshold voltage of −13 V. Also, TiN-gated devices provide improved and better breakdown performance as compared to Pt-gated devices. These results show that breakdown is determined by the combined effect of gate metal, electric-field distribution, and surface leakage. TiN and Pt gate contacts can be selected to balance current-drive and threshold-voltage requirements while enabling an Au-free gate-stack approach that is more compatible with industrial CMOS fabrication practices for Al-rich AlGaN RF- and power electronics.

**Acknowledgment**

**Distribution Statement A: Approved for public release. Distribution is unlimited.**

This project is funded by the Microelectronics Commons Program, a DoW initiative. This material is based upon work supported by ADI under award number 10015113. Our work is also supported by the ONR grant number N000142312010. (Capt. Lynn Petersen). The reported work is also partially supported by the Air Force Office of Scientific Research under award number FA9550-24-1-0269. Any opinions, findings, and conclusions or recommendations expressed in this material are those of the author(s) and do not necessarily reflect the views of the United States Air Force (Dr. Ali Sayir). This work is also supported by the USC Research Institute for Extreme Electronics, USC grant number 80005646.

## AUTHOR DECLARATIONS

### Conflict of Interest

The authors have no conflicts to disclose.

## DATA AVAILABILITY

The data that support the findings of this study are available from the corresponding author upon reasonable request**.**

**Figure 1:**

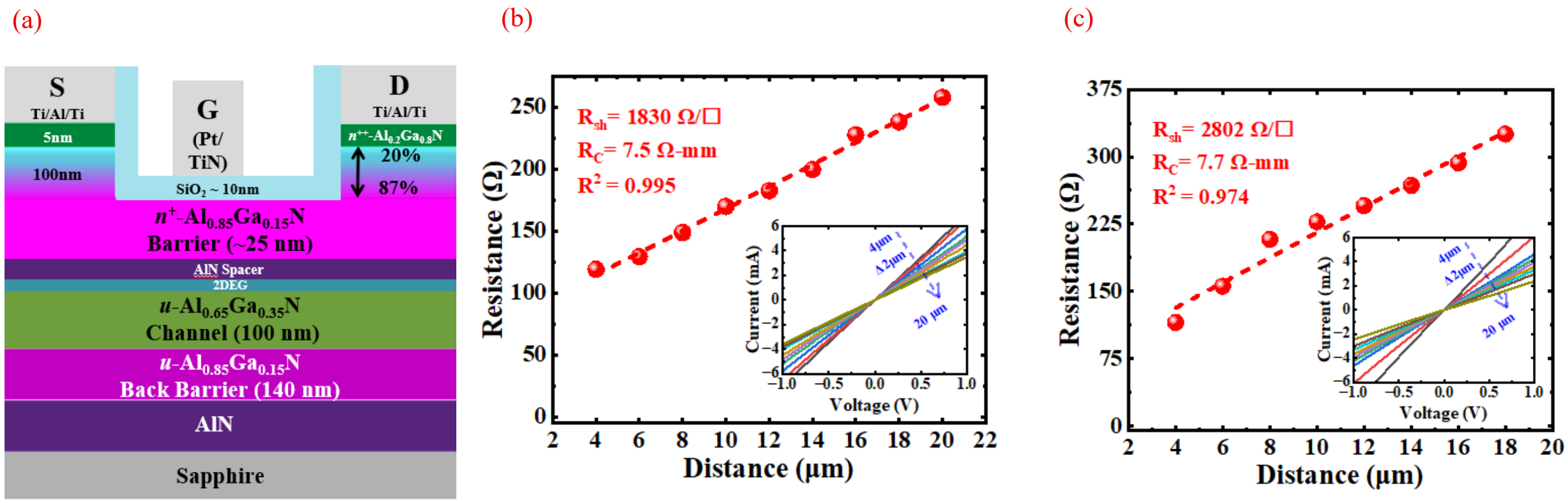

**Fig 1: (a) Schematic illustration of the epitaxial and fabricated device structure for MOS-HEMT, (b and c) computed $R_c$ and $R_{sh}$ from linear fitting of the measured data: (b) electron-beam Ti/Al/Ti contacts, (c) sputtered Ti/Al/Ti contacts. (Inset shows the I-V for different TLM spacings).**

**Table I:** **Description of the Samples**

| Contact Platform | Device | Ohmic Contact | Gate Stack | $I_D$ (mA/mm) | $V_{th}$ (V) | $R_{on}$ (Ω·mm) |
|---|---|---|---|---|---|---|
| Au-free (this work) | **A** | Ti/Al/Ti, e-beam | TiN, sputtered | ~ 450 | −13.0 ± 0.5 | ~ 28 |
| | **B** | Ti/Al/Ti, e-beam | Pt, sputtered | ~ 283 | −8.0 ± 0.3 | ~ 33 |
| | **C** | Ti/Al/Ti, sputtered | TiN, sputtered | ~ 278 | −10.0 ± 0.8 | ~ 49 |
| | **D** | Ti/Al/Ti, sputtered | Pt, sputtered | ~ 220 | −3.0 ± 0.4 | ~ 41 |
| Au-based [1] | **Reference** | Ti/Al/Ti/Au, e-beam | Ni/Au, e-beam | ~ 410 | −17.3 ± 0.3 | ~ 24 |

**Figure 2:**

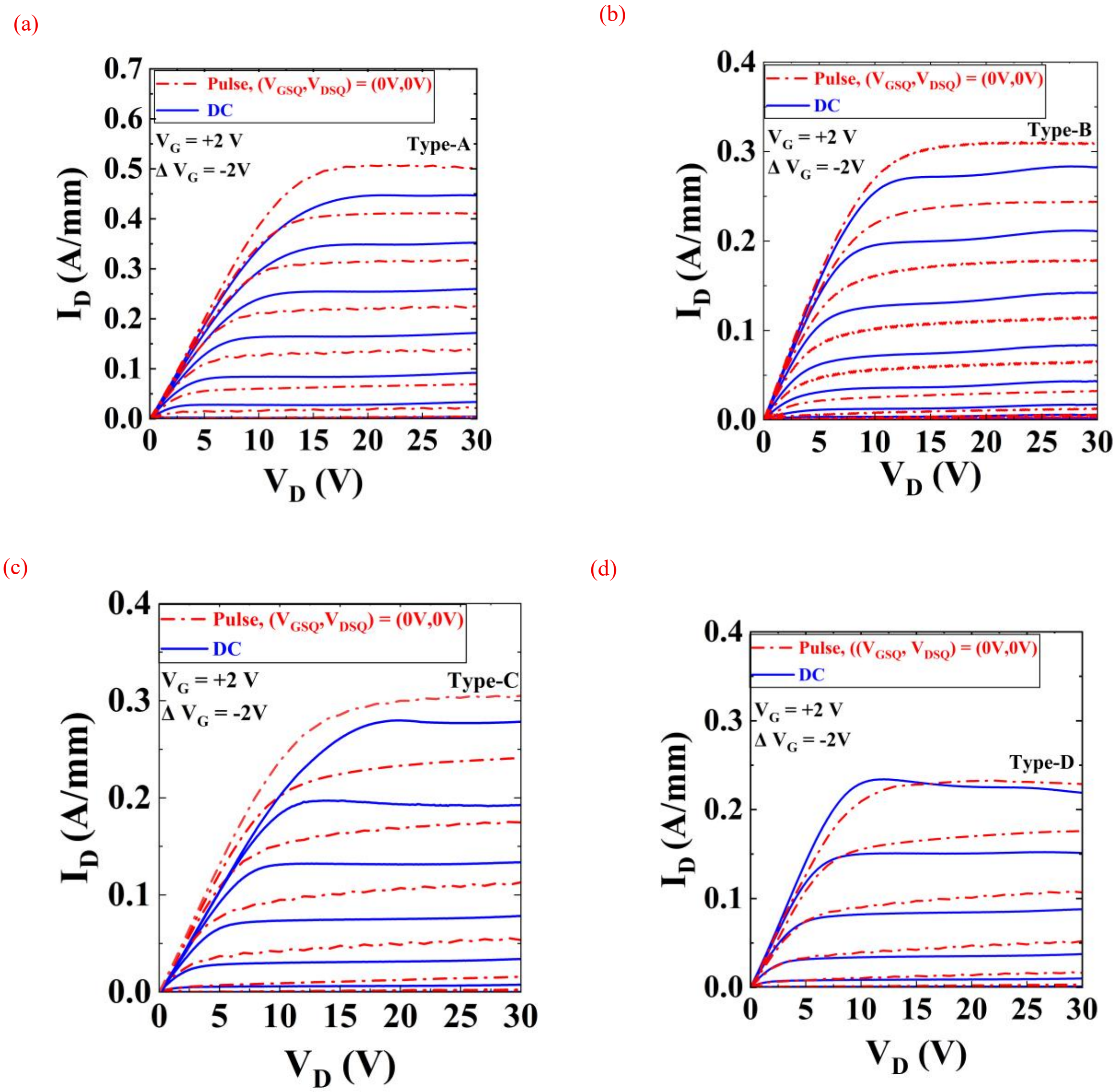

**Fig 2: Measured output characteristics of DC and Pulse at (0V,0V) (a) TiN gate – electron-beam ohmic contact, (b) Pt gate – electron-beam ohmic contact, (c) TiN gate – sputtered ohmic contact, (d) Pt gate – sputtered ohmic contact.**

**Figure 3:**

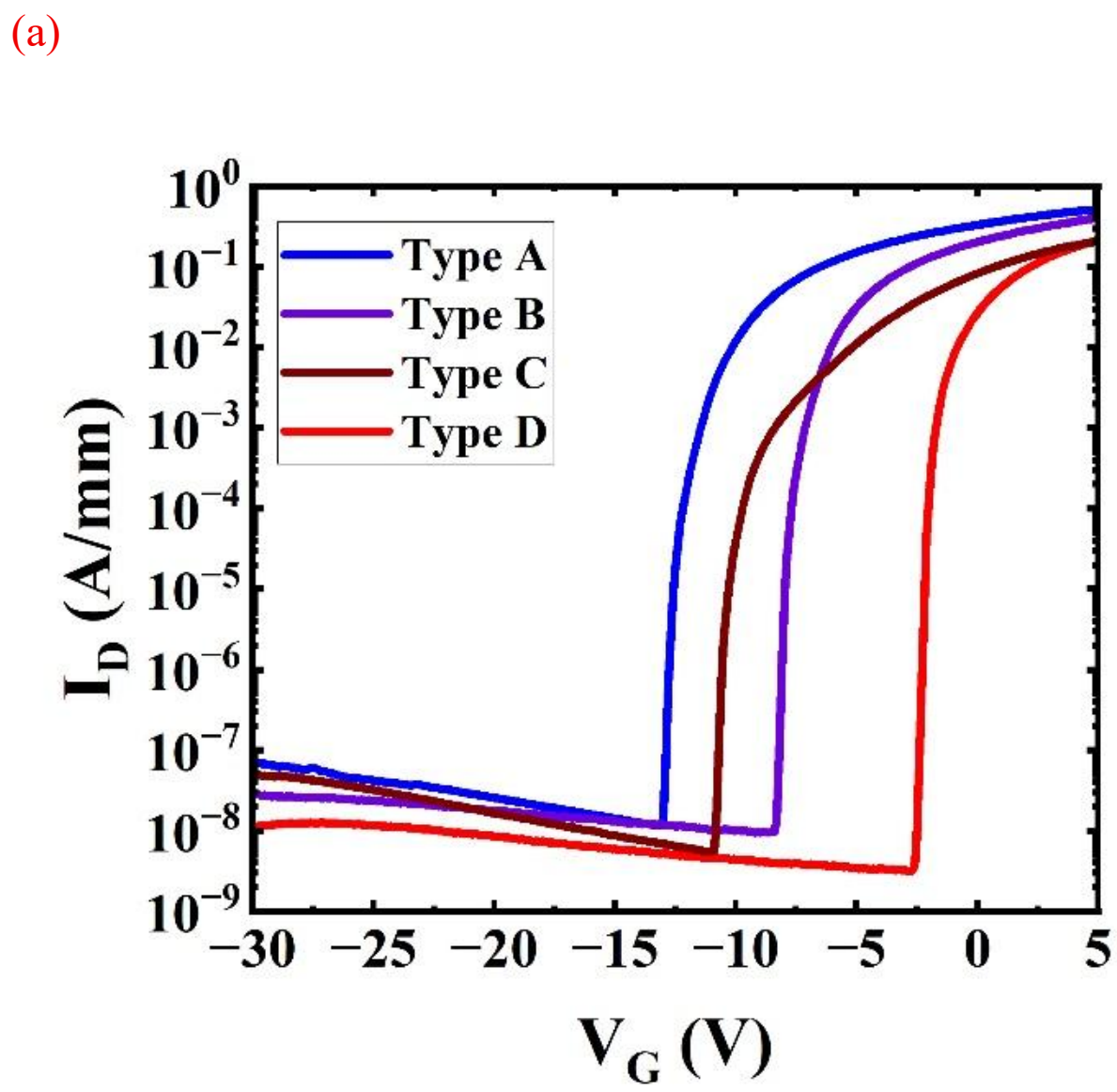

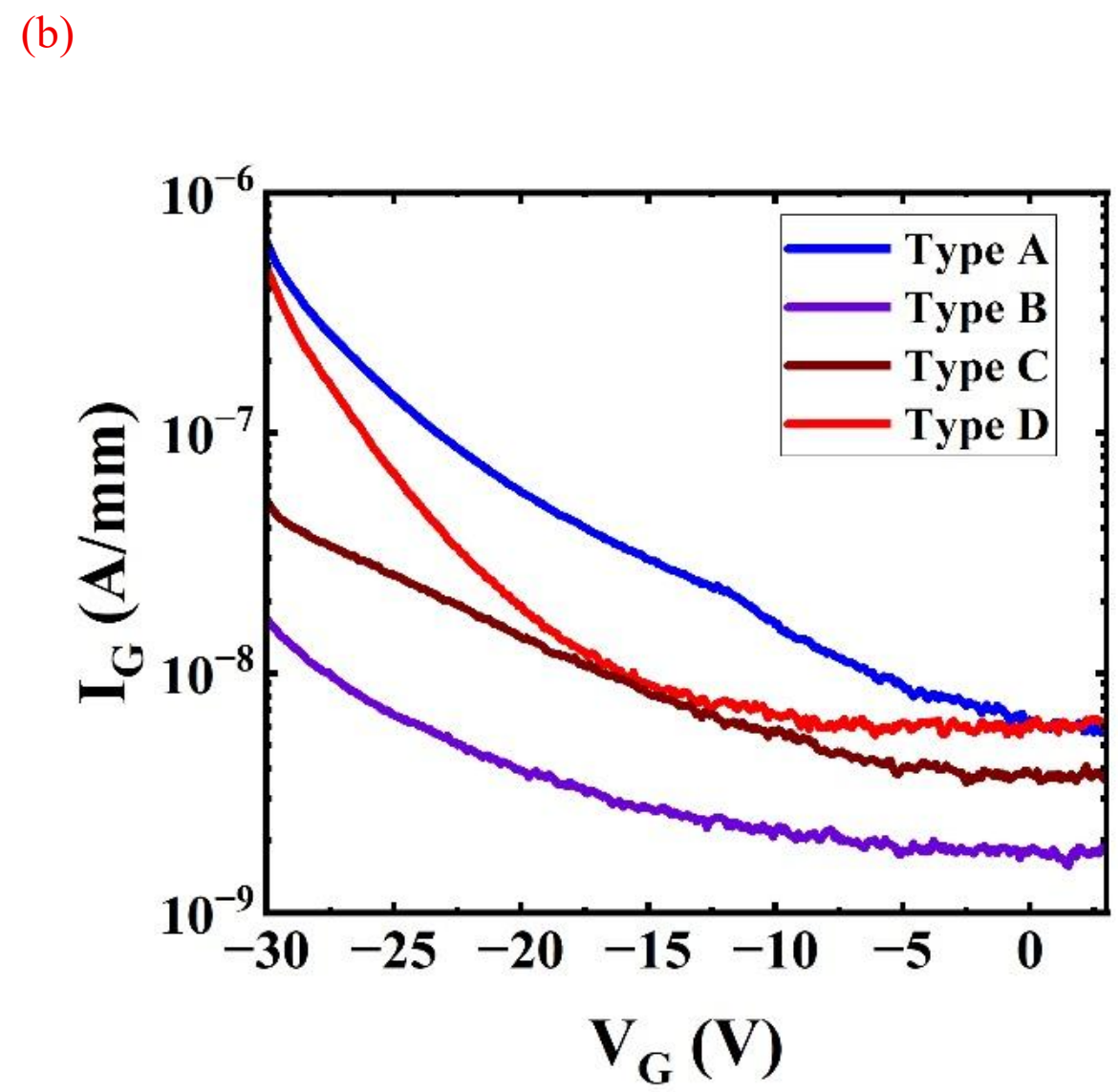

**Fig 3: Measured electrical characteristics of each device type (A, B, C, D): (a) transfer characteristics ($I_D$–$V_G$) plotted on a logarithmic scale and (b) gate-leakage characteristics ($I_G$-$V_G$) measured at $V_D$=20V.**

**Figure 4:**

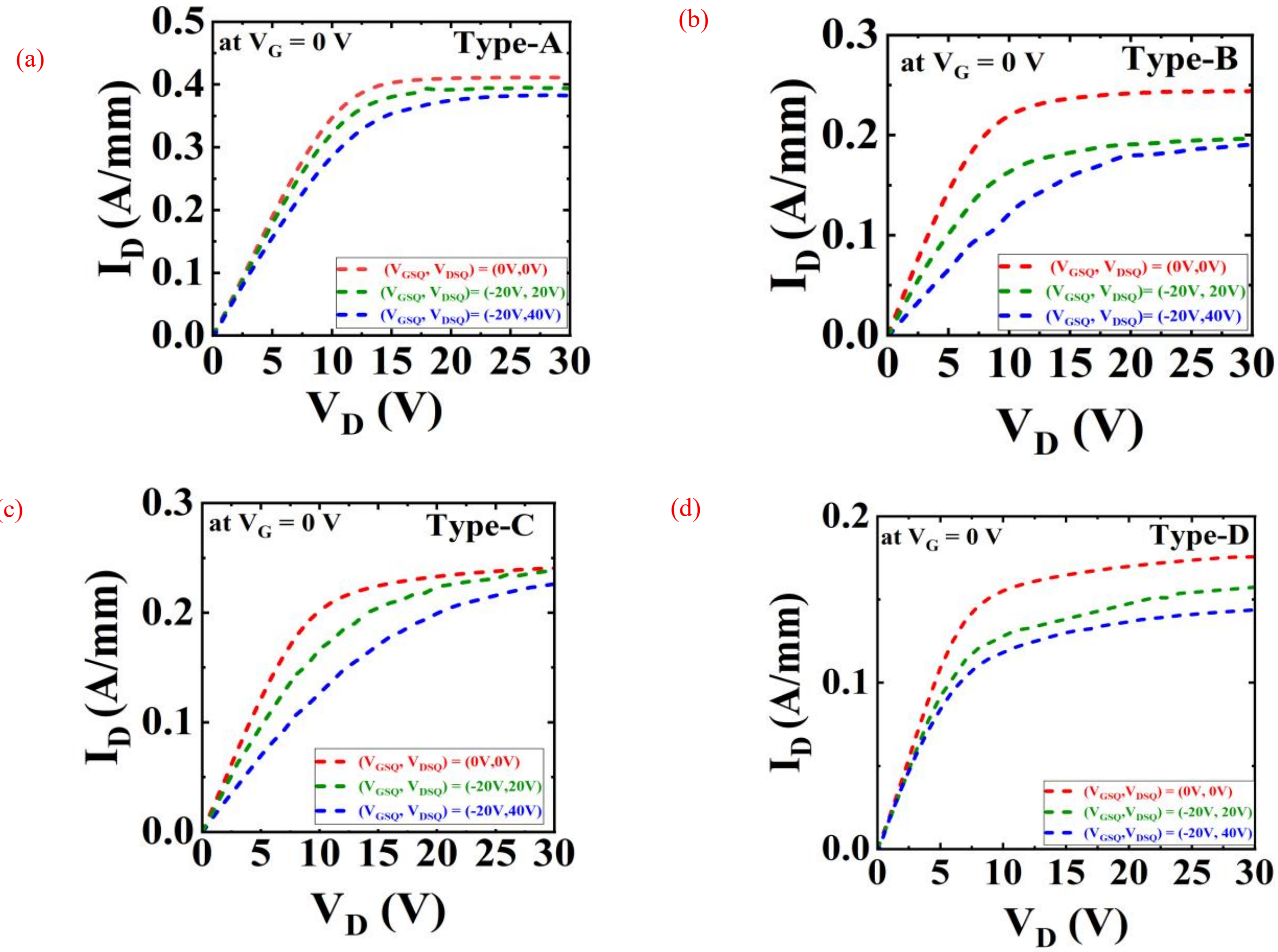

**Fig 4: Measured Pulsed I-V characteristics at a pulse width of 6 μs and 1 ms off-time (duty cycle ≈ 0.6%) under different quiescent bias conditions (a) TiN gate – electron-beam ohmic contact, (b) Pt gate – electron-beam ohmic contact, (c) TiN gate – sputtered ohmic contact, (d) Pt gate – sputtered ohmic contact.**

**Figure 5:**

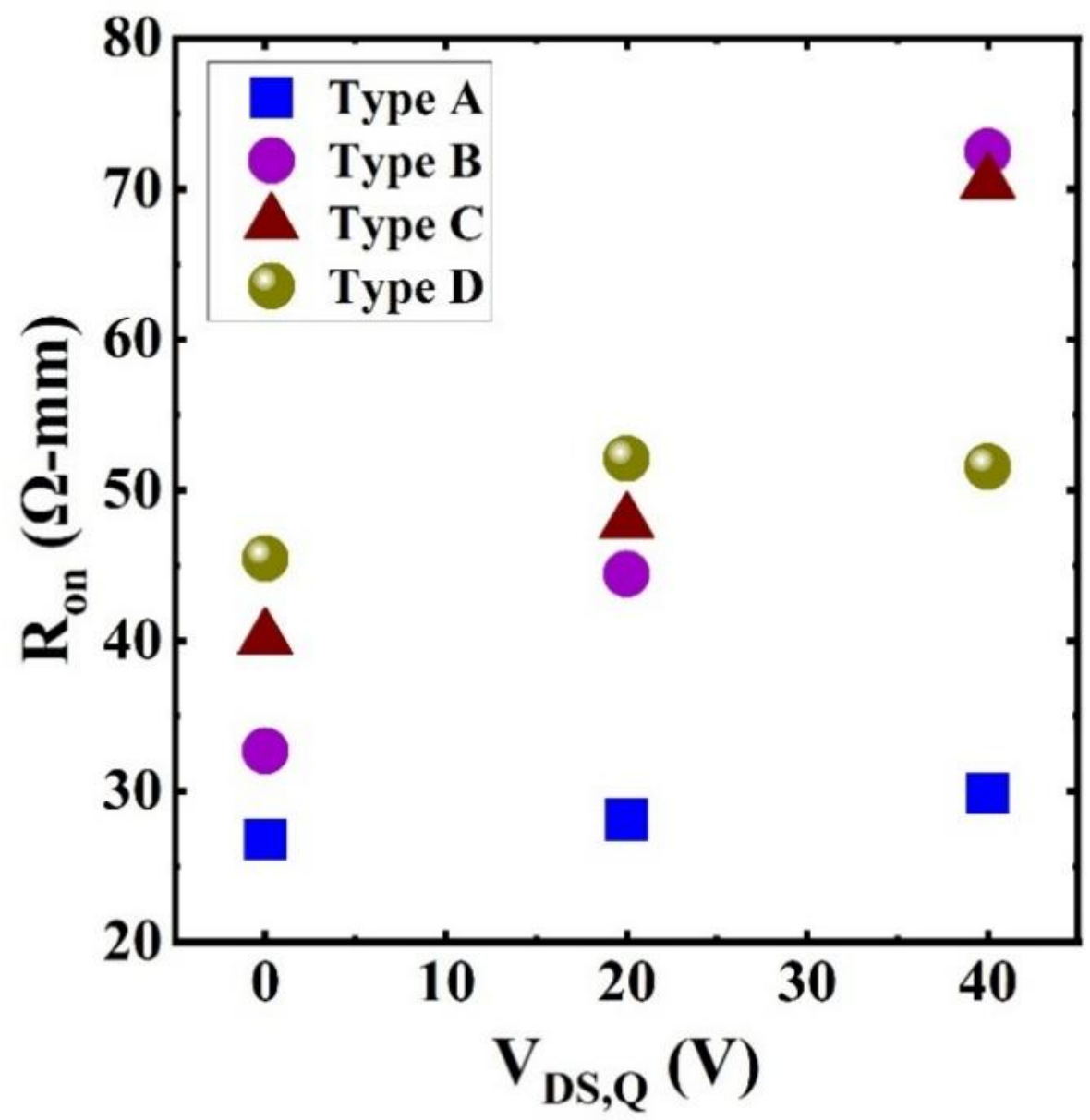

**Fig 5: Measured Dynamic $R_{on}$ of each device type extracted under quiescent bias conditions of ($V_{GS,Q}$, $V_{DS,Q}$) = (0V, 0V), (-20V, 20V), and (-20V, 40V) at 0.6% Duty Cycle.**

**Figure 6:**

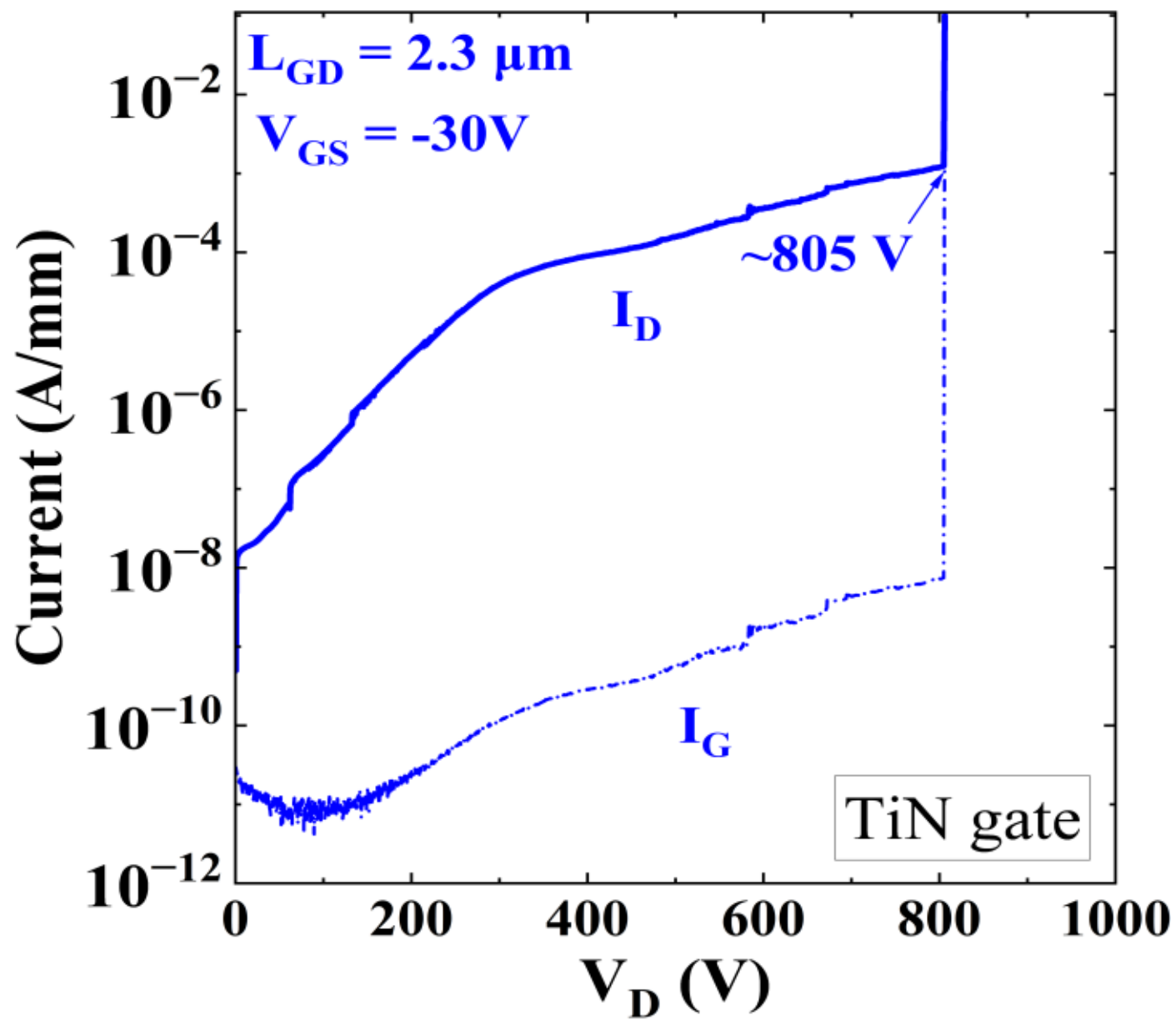

**Fig 6: Measured Off-state breakdown characteristics for TiN-gated device with $L_{GD}$ = 2.3 µm. The solid and dashed lines are the $I_D$-$V_D$ and $I_G$-$V_D$ curves, respectively, at $V_{GS}$ = -30V.**

**Figure 7:**

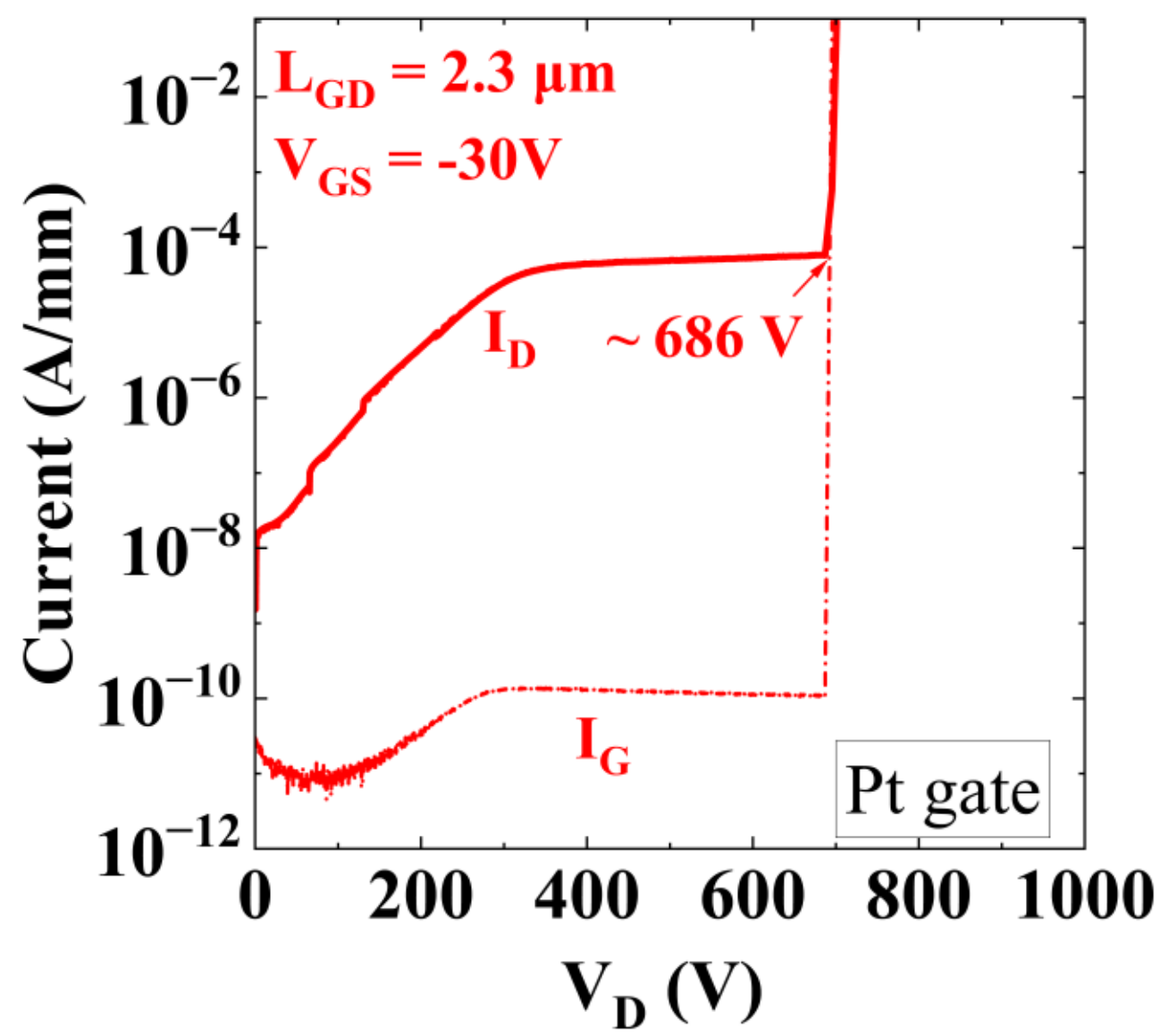

**Fig 7: Measured Off-state breakdown characteristics for Pt-gated device with $L_{GD}$ = 2.3 $\mu$m. The solid and dashed lines are the $I_D$-$V_D$ and $I_G$-$V_D$ curves, respectively, at $V_{GS}$ = -30V.**

**Figure 8:**

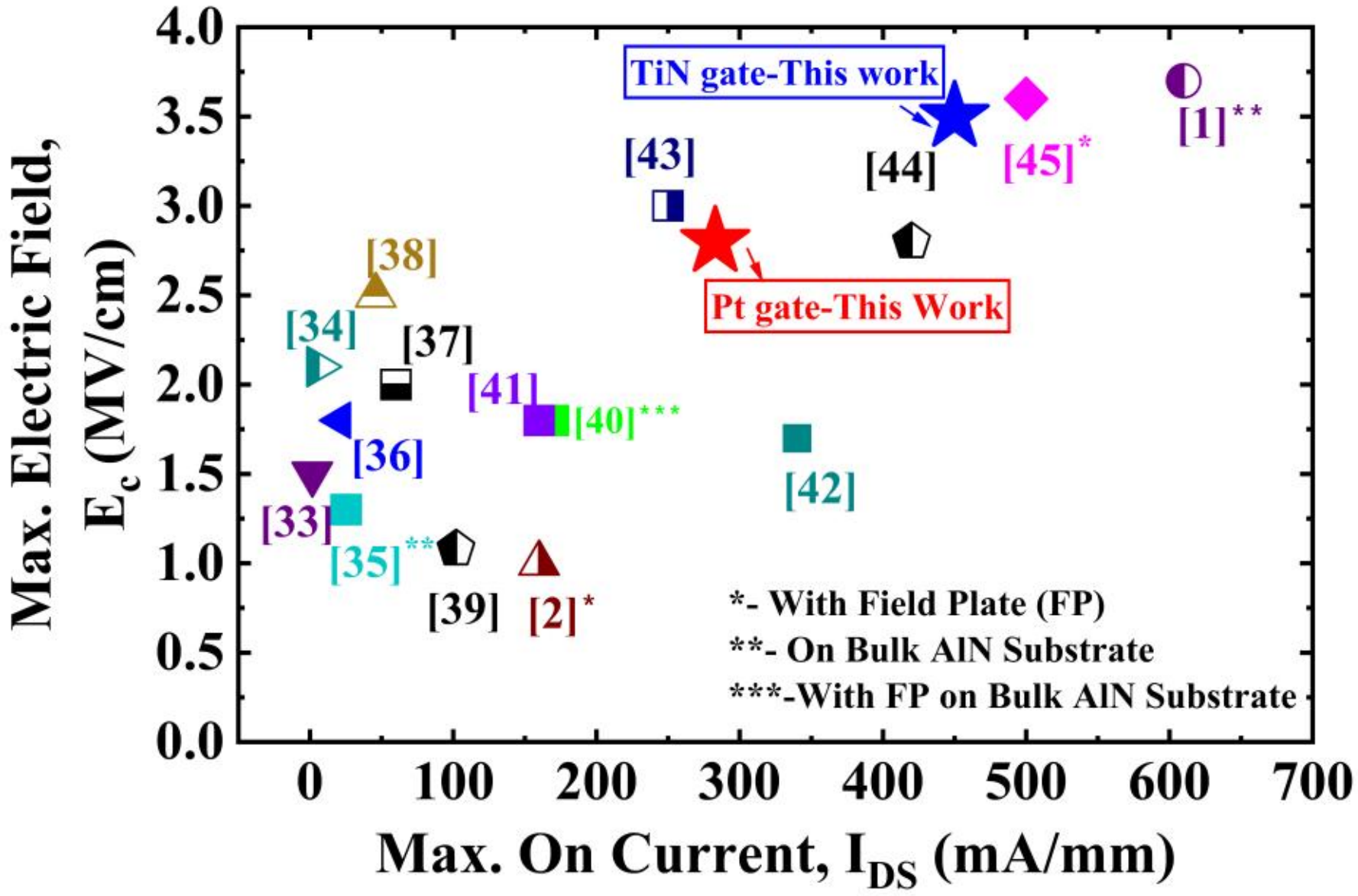

**Fig 8: Benchmark comparison of maximum drain current density ($I_{DS,max}$) and average electric-field ($E_c$) Al-rich AlGaN HEMTs reported in literature, including the TiN and Pt-gated device in this work.**